\documentclass[journal]{IEEEtran}
\ifCLASSINFOpdf
  \usepackage[pdftex]{graphicx}
\else
\fi
\usepackage{cite}
\usepackage{amsmath,amssymb,amsfonts}
\usepackage{algorithmic}
\usepackage{graphicx}
\usepackage{textcomp}
\usepackage{xcolor}
\usepackage{amsmath}
\usepackage{algorithmic}
\usepackage{array}
\usepackage{stfloats}
\usepackage{array}
\begin{document}
\title{Burst-Mode Digital Signal Processing for Coherent Optical Time-Division Multiple Access}

\author{Ji Zhou, Cheng Li, Haide Wang, Zhiyang Liu, Weiping Liu, and  Changyuan Yu
\thanks{Manuscript received xxxx; revised xxxx. This work was supported in part by the National Key R\&D Program of China under Grant 2023YFB2905700, in part by the National Natural Science Foundation of China under Grant 62371207 and Grant 62005102, in part by the Young Elite Scientists Sponsorship Program by CAST under Grant 2023QNRC001, and in part by the Hong Kong Research Grants Council GRF under Grant 15231923. (\it{Corresponding author: Haide Wang.})}
\thanks{Ji Zhou, Zhiyang Liu, and Weiping Liu are with Department of Electronic Engineering, College of Information Science and Technology, Jinan University, Guangzhou 510632, China.}
\thanks{Cheng Li and Changyuan Yu are with Department of Electrical and Electronic Engineering, The Hong Kong Polytechnic University, Hong Kong.}
\thanks{Haide Wang is with School of Cyber Security, Guangdong Polytechnic Normal University, Guangzhou 510665, China.}
}

\maketitle
\begin{abstract} \boldmath
As the 50G optical access gradually matures, it is time to discuss Beyond 50G optical access. According to the evolution rules of optical access standards, Beyond 50G optical access data rate may achieve 200Gb/s. Direct detection faces great challenges for Beyond 50G optical access, which makes coherent detection a potential solution. Similar to 50G optical timing-division-multiple access (TDMA), burst-mode digital signal processing (BM-DSP) is also required for Beyond 50G coherent optical TDMA (CO-TDMA). This paper proposes coherent BM-DSP (Co-BM-DSP) based on approximately 10ns designed preambles to process the burst signal for 200G CO-TDMA, which can fast estimate the state of polarization, frequency offset, sampling phase offset, synchronization position, and equalizer coefficients. Meanwhile, for obtaining the equalizer coefficients based on the designed preamble, the channel estimation based on the minimum-mean-square-error criterion is theoretically proven to have a unique solution for ensuring reliability. In conclusion, the proposed Co-BM-DSP based on the designed preambles paves the way for the applications of Beyond 50G CO-TDMA.
\end{abstract}

\IEEEpeerreviewmaketitle

\section{Introduction}
\IEEEPARstart{I}N 2021, significant progress was made in the International Telecommunication Union Telecommunication (ITU-T) standardization sector to define a higher-speed (HS) optical access with a line rate of 50Gb/s \cite{ITUT202150gigabit,9743347, zhang2020progress}. Recently, 50G optical access has gradually matured with the efforts of vendors and operators. It is time to discuss and research Beyond 50G optical access. ITU-T establishes a group for discussing future optical-access-related technologies for Beyond 50G optical access\cite{ITUT2023b50gigabit, luo2024beyond, nesset2023next}. There are two views on the line rate of Beyond 50G optical access: 100Gb/s or 200Gb/s. Based on the evolution rules of ITU-T standards, the data rate of optical access is 4 or 5 times higher than that of previous-generation optical access. Thus, Beyond 50G optical access data rate may achieve 200Gb/s \cite{10023536, 10493050, wei2024physical}. For the 200G optical access, direct detection (DD) faces tough challenges. Coherent detection becomes the most potential solution \cite{10403903, zhang2022coherent, zhang2023flexible}.

\begin{figure}[!t]
    \centering
    \includegraphics[width=\linewidth]{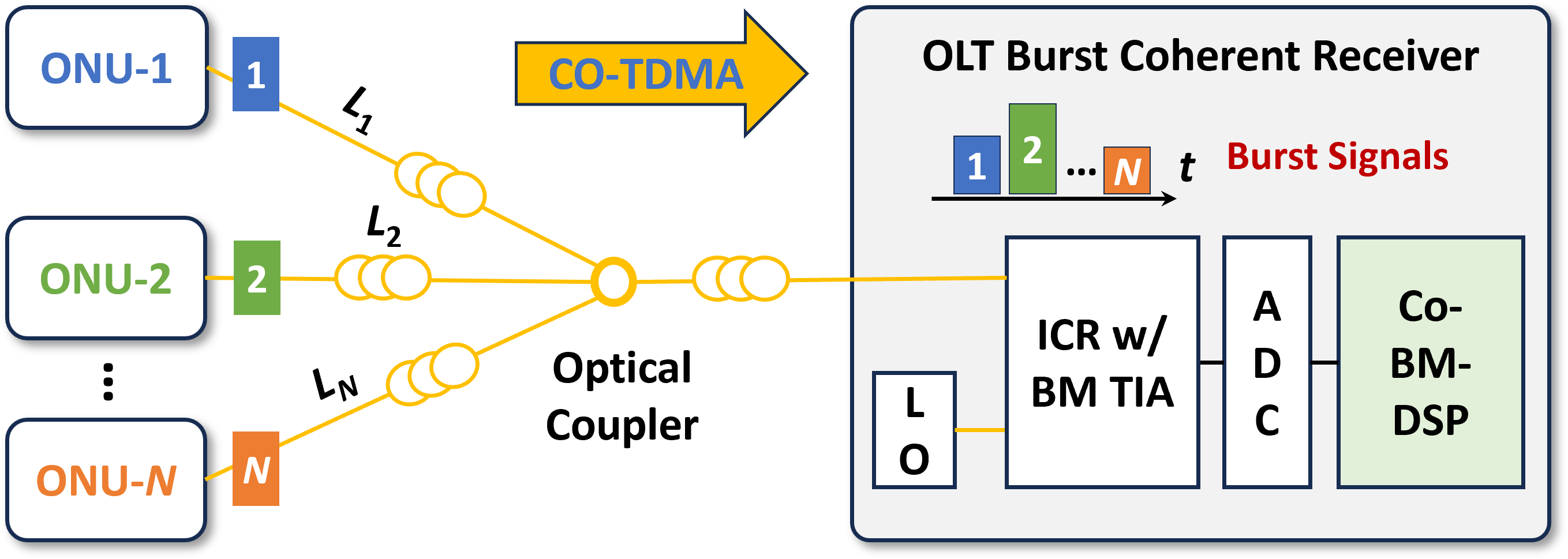}
    \caption{(a) Schematic diagram of CO-TDMA based on Co-BM-DSP. ONU: optical network unit, OLT: optical line terminal, LO: local oscillator, ICR: integrated coherent receiver, ADC: analog-to-digital converter.}
    \label{TDMA}
\end{figure}

Owing to statistical multiplexing, time division multiple access (TDMA) has been used to ensure capacity and quantity for the subscribers since the beginning of commercial applications for the optical access \cite{chung2022tdm,li2020dsp, van2024benefits}. In 50G DD optical TDMA, one burst-mode trans-impedance amplifier (BM-TIA) and burst-mode digital signal processing (BM-DSP) are required to receive and process the upstream burst signal, which has a short convergence time to improve spectral efficiency \cite{10484954, 10209815,yang2023fpga}. Similarly, coherent optical TDMA (CO-TDMA) shown in Fig. \ref{TDMA} needs four BM-TIAs and coherent BM-DSP (Co-BM-DSP) to process the burst polarization-division-multiplexing (PDM) in-phase and quadrature (IQ) signals such as PDM quadrature phase shift keying and 16-quadrature amplitude modulation (16QAM) signal \cite{9296729, 10117057, li2023efficient}. The difficulty of the Co-BM-DSP is that the state of polarization (SOP) and frequency offset cause estimation errors in the sampling phase offset (SPO), synchronization position, and tap coefficients of the equalizer. Therefore, except for the SPO, synchronization position, and tap coefficients, the Co-BM-DSP should quickly estimate the SOP and frequency offset \cite{8513878, koma2016demonstration, xing2022demonstration}. 

In this paper, we design 10ns preambles with 320 symbols for the Co-BM-DSP in 200G CO-TDMA based on a 32GBaud PDM-16QAM signal. The two frequency tones with 128 symbols on X polarization and Y polarization are used to estimate SOP, frequency offset, and SPO. The 3$\times$64 constant amplitude zero auto-correlation (CAZAC) symbols are used to obtain the synchronization position and equalizer coefficients. This paper is an extended version of our previous work \cite{Wang2024} published on ECOC 2024, which provides more details about principle of Co-BM-DSP, discussion of the experimental results, and reliability analysis for channel estimation. The main contributions of this paper are as follows: 
\begin{itemize}
\item We propose Co-BM-DSP based on the designed preambles to estimate SOP, frequency offset, SPO, synchronization position, and equalizer coefficients to process the burst signal for CO-TDMA.
\item The feasibility of the proposed Co-BM-DSP based on the 10ns designed preambles with 320 symbols is experimentally verified by 200G CO-TDMA based on a 32GBaud PDM-16QAM signal.
\item The channel estimation based on the minimum-mean-square-error (MMSE) criterion and the designed 3$\times$64 CAZAC symbols is theoretically proven to have a unique solution for ensuring reliability.
\end{itemize}

The remainder of this paper is organized as follows. The structure of the designed preambles and the principle of the Co-BM-DSP for CO-TDMA are shown in Section \ref{SectionII}. In Section \ref{SectionIII}, the experimental setups of the 200G CO-TDMA are introduced. In Section \ref{SectionIV}, the experimental results are discussed. The paper is concluded in Section \ref{SectionV}. Finally, a reliability analysis for channel estimation based on the MMSE algorithm is given in Appendix \ref{AppendicesA}.

\section{Designed Preambles and Co-BM-DSP}\label{SectionII}
This section introduces the structure of the designed preambles and the Co-BM-DSP principle for CO-TDMA.

\begin{figure}[!t]
    \centering
    \includegraphics[width=\linewidth]{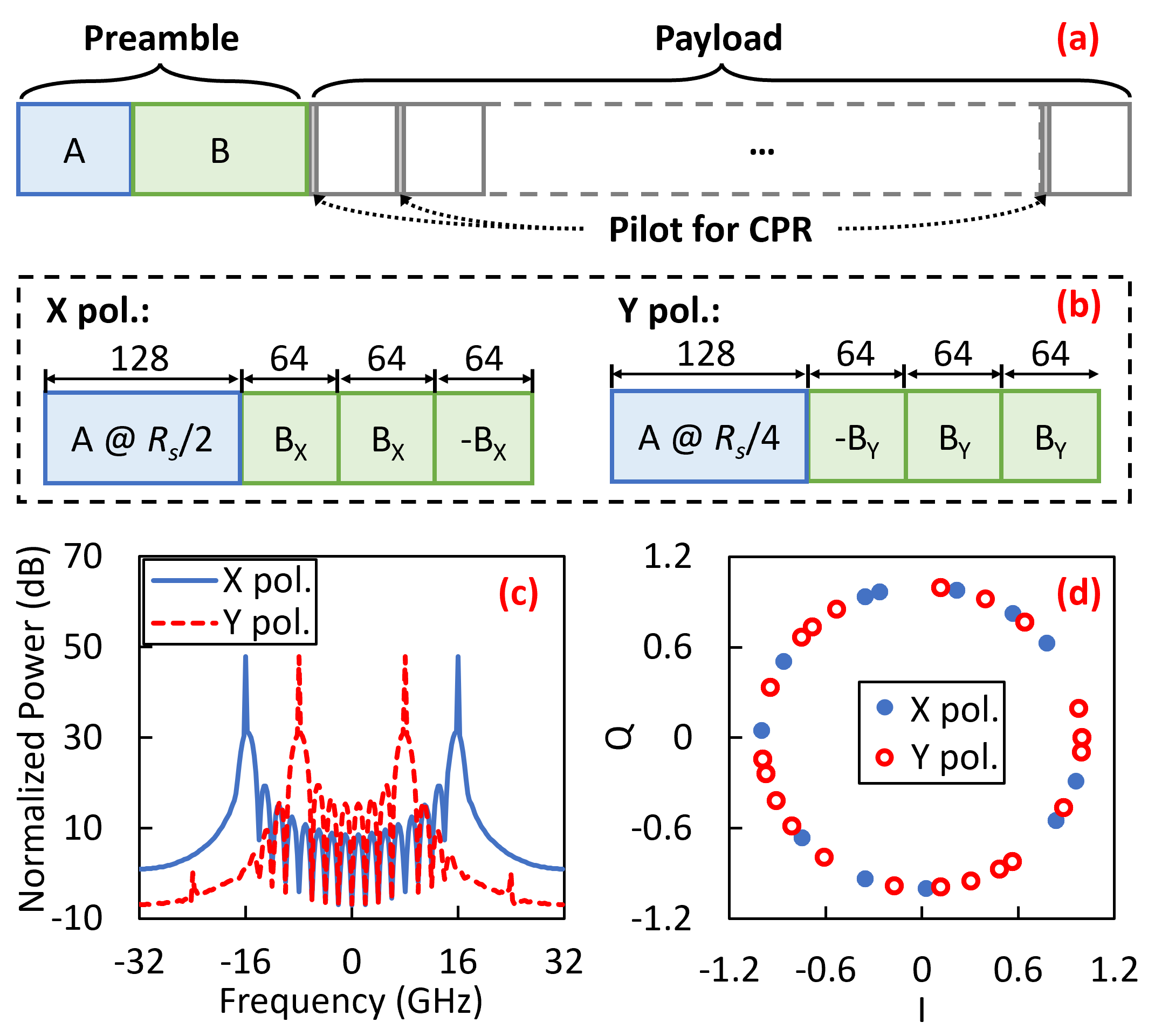}
    \caption{(a) Frame structure for burst signal of CO-TDMA. (b) Structure of designed preambles including Preambles A and B. (c) The spectra of Preamble A for X polarization and Y polarization with two different frequency tones. (d) Constellation diagrams of the Preamble B for X polarization and Y polarization with two different seeds.}
    \label{Preamble}
\end{figure}

\subsection{Designed Preambles for CO-TDMA}
Figure \ref{Preamble}(a) shows the frame structure for the burst signal of CO-TDMA. Before the payload symbols, the designed preambles consist of two parts: Preamble A and Preamble B. In addition, one pilot symbol is inserted into every 32 payload symbols for the pilot-based carrier phase recovery (CPR). Fig. \ref{Preamble}(b) depicts the structure of designed preambles including Preambles A and B. Preamble A with 128 symbols is designed as two frequency tones at half of the baud rate (i.e. $R_\text{s}/2$ where $R_\text{s}$ denotes the baud rate) and a quarter of the baud rate (i.e. $R_\text{s}/4$) for the X polarization and Y polarization, respectively. Preamble B uses three sets of 64 CAZAC symbols for estimating equalizer coefficients and the synchronization position. The three sets of 64 CAZAC symbols are multiplied by the coefficients of $1$, $1$, and $-1$ for the X polarization. For the Y polarization, the coefficients are $-1$, $1$, and $1$. Fig. \ref{Preamble}(c) shows the spectra of Preamble A where the frequency tones are located at 8GHz and 16GHz for X polarization and Y polarization when the $R_\text{s}$ is set to 32GBaud, respectively. Fig. \ref{Preamble}(d) shows constellation diagrams of the Preambles $\mathbf{B}_{\mathrm{X}}$ and $\mathbf{B}_{\mathrm{Y}}$ with different seeds. Preambles $\mathbf{B}_{\mathrm{X}}$ and $\mathbf{B}_{\mathrm{Y}}$ have ideal auto-correlation, which are appropriate for the frame synchronization. Meanwhile, their time-domain and frequency-domain envelopes are constant, making Preamble B suitable for feed-forward channel estimation.

\begin{figure}[!t]
    \centering
    \includegraphics[width=\linewidth]{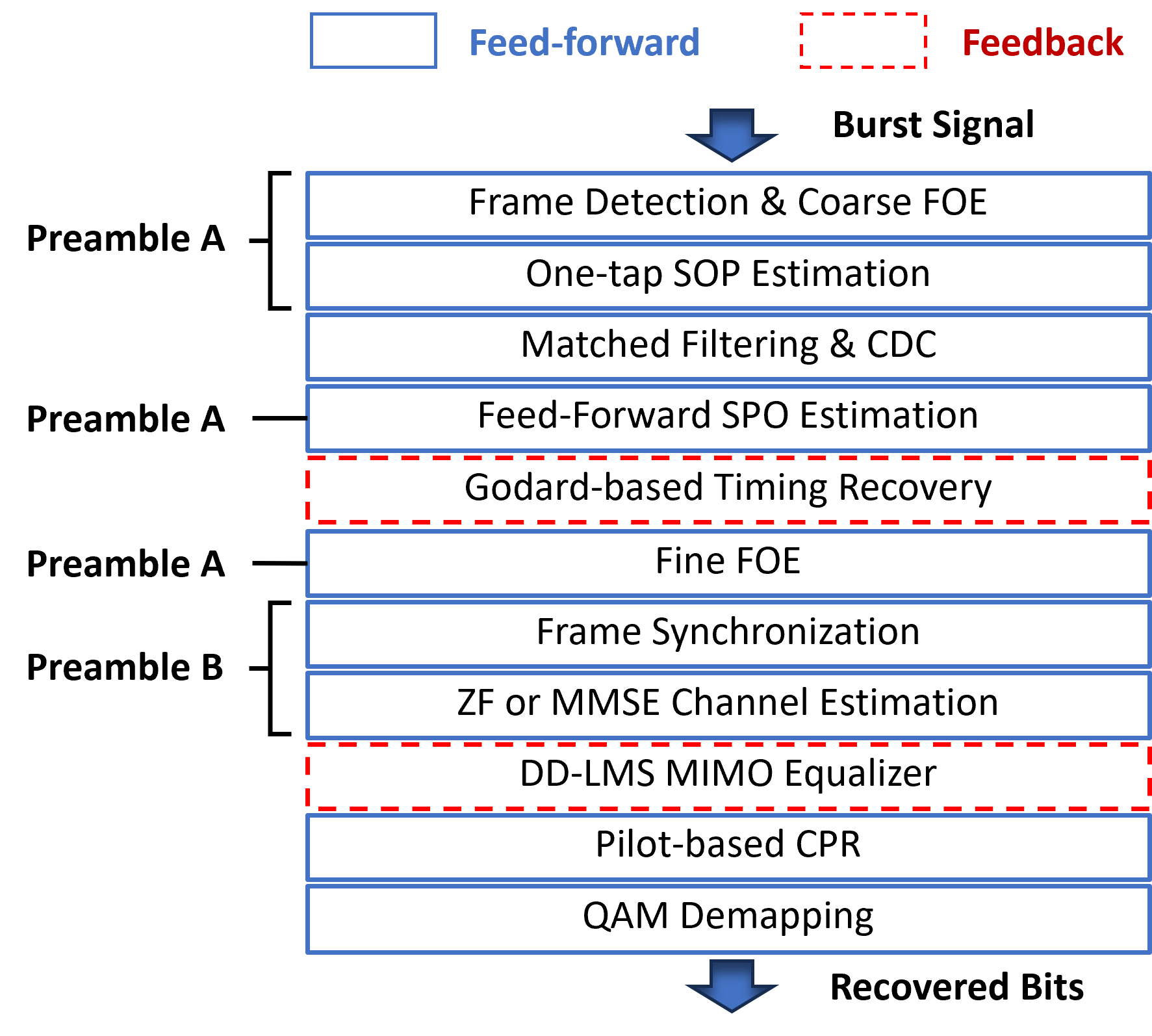}
    \caption{Block diagram of Co-BM-DSP based on the designed preambles at the receiver side. FOE: frequency offset estimation, CDC: chromatic dispersion compensation, ZF: zero forcing, DD-LMS: decision-direct least-mean-square, MIMO: multiple-input-multiple-output, CPR: carrier phase recovery.}
    \label{DSP}
\end{figure}

\subsection{Co-BM-DSP Based on Designed Preambles}
Figure \ref{DSP} shows the block diagram of Co-BM-DSP based on the designed preambles at the receiver side. Based on Preamble A, frame detection, coarse frequency offset estimation (FOE), and one-tap SOP estimation can be implemented. After the matched filtering and chromatic dispersion compensation (CDC), the feed-forward SPO estimation and fine FOE are implemented by Preamble A to initialize the Godard-based timing recovery and finely compensate frequency offset, respectively. Based on Preamble B, frame synchronization and feed-forward channel estimation can be realized. MMSE and zero-forcing (ZF) algorithms can implement the feed-forward channel estimation to calculate the tap coefficients. After initializing the tap coefficients calculated by the proposed channel estimation, the tap coefficients of 2$\times$2 multiple-input-multiple-output (MIMO) equalizer are updated by the feedback decision-direct least mean square (DD-LMS) algorithm. Finally, after the pilot-based CPR, the recovered 16QAM is de-mapped into bits. Next, the Co-BM-DSP algorithms will be introduced in detail.

\subsubsection{Frame Detection and Coarse FOE Based on Preamble A}
The received burst signal is first sliced within a sliding window to detect the burst frame. Then the sliced signals are transferred to the frequency domain to search two symmetric frequency tones of Preamble A. Once two symmetric frequency tones are recognized, the arrival time of one burst frame is confirmed. After the frame detection, the frequency offset should be estimated and compensated to ensure the signal spectrum within the passband of the matched filter. Based on Preamble A, the FOE can be implemented by obtaining the center frequency offset between two symmetric tones. However, discrete Fourier transform (DFT) size determines the frequency resolution, limiting the accuracy of FOE based on Preamble A. Therefore, the FOE based on Preamble A is still coarse, and a finer FOE is required. More details about the frame detection and coarse FOE based on Preamble A can be found in References \cite{10042001, wang2023non}.

\subsubsection{One-Tap SOP Estimation and Recovery Based on Preamble A}
The frequency-domain Preamble A of the X polarization or Y polarization is the tone $T_{\mathrm{X}, R_\text{s}/2}$ at half of the baud rate $R_\text{s}/2$ or the tone $T_{\mathrm{Y}, R_\text{s}/4}$ at a quarter of the baud rate $R_\text{s}/4$, respectively. According to the Jones matrix \cite{liu2009initial, savory2010digital}, the received Preamble A of the X polarization and Y polarization can be modeled as
\begin{equation} \label{Eq1}
\begin{aligned}
&{\left[\begin{array}{ll}
R_{\mathrm{X}, R_\text{s} / 2} & R_{\mathrm{X}, R_\text{s} / 4} \\
R_{\mathrm{Y}, R_\text{s} / 2} & R_{\mathrm{Y}, R_\text{s} / 4}
\end{array}\right]} =\\
&\left[\begin{array}{cc}
\sqrt{1-\alpha} \cdot T_{\mathrm{X}, R_\text{s} / 2} & -\sqrt{\alpha} e^{j \theta} \cdot T_{\mathrm{Y}, R_\text{s} / 4} \\
\sqrt{\alpha} e^{-j \theta} \cdot T_{\mathrm{X}, R_\text{s} / 2} & \sqrt{1-\alpha} \cdot T_{\mathrm{Y}, R_\text{s} / 4}
\end{array}\right] 
\end{aligned}
\end{equation}
where $\alpha$ is the power split ratio and $\theta$ is the relative phase difference between the X polarization and Y polarization \cite{kikuchi2008polarization, koma2018burst}. The $\alpha$ can be calculated by
\begin{equation}
\frac{\left|R_{\mathrm{X}, R_\text{s} / 2}\right|^2+\left|R_{\mathrm{Y}, R_\text{s} / 4}\right|^2}{\left|R_{\mathrm{Y}, R_\text{s} / 2}\right|^2+\left|R_{\mathrm{X}, R_\text{s} / 4}\right|^2}=\frac{1-\alpha}{\alpha}
\end{equation}
where $\left|\cdot\right|^2$ denotes the power of the tones. After SOP recovery by multiplexing the inverse Jones matrix, Preamble A can be expressed as
\begin{equation}
\begin{aligned}
& {\left[\begin{array}{ll}
S_{\mathrm{X}, R_\text{s} / 2} & S_{\mathrm{X}, R_\text{s} / 4} \\
S_{\mathrm{Y}, R_\text{s} / 2} & S_{\mathrm{Y}, R_\text{s} / 4}
\end{array}\right] }= \\
&\left[\begin{array}{cc}
\sqrt{1-\alpha} & \sqrt{\alpha} e^{-j \theta} \\
-\sqrt{\alpha} e^{j \theta} & \sqrt{1-\alpha}
\end{array}\right]\times \left[\begin{array}{ll}
R_{\mathrm{X}, R_\text{s} / 2} & R_{\mathrm{X}, R_\text{s} / 4} \\
R_{\mathrm{Y}, R_\text{s} / 2} & R_{\mathrm{Y}, R_\text{s} / 4}
\end{array}\right].
\end{aligned}
\end{equation}
When the $\alpha$ is confirmed, $\theta$ can be searched by maximizing the power sum of $S_{\mathrm{X}, R_\text{s} / 2}$ of the X polarization and $S_{\mathrm{Y}, R_\text{s} / 4}$ of the Y polarization, which can be calculated by
\begin{equation}
\begin{aligned}
\theta_\text{max} = &\underset{\theta}{\operatorname{argmax}}\left(\left|S_{\mathrm{X}, R_\text{s} / 2}\right|^2+\left|S_{\mathrm{Y}, R_\text{s} / 4}\right|^2\right) 
\end{aligned}
\end{equation}
where ${\operatorname{argmax}}(\cdot)$ represents the set that maximizes the input variable and $0 \leq \theta < 2\pi$.

\subsubsection{Phase-Initialized Timing Recovery Based on Preamble A}  \label{SecTR}
After the one-tap SOP estimation and recovery, the recovered Preamble A $S_{\mathrm{X}, R_\text{s} / 2}$ and $S_{\mathrm{Y}, R_\text{s} / 4}$ can be used for the feed-forward SPO estimation for X polarization and Y polarization, respectively. Using the tone at $R_\text{s}/2$, the initial SPO $\tau_{0, R_\text{s} / 2}$ can be estimated by
\begin{equation}
\tau_{0, R_\text{s} / 2}=\frac{K}{2\pi} \arg \left[S(N/2 - N/K/2) S^*(N/2 + N/K/2)\right]
\end{equation}
where $K$ is the oversampling rate and $N$ is the DFT size. $\arg(\cdot)$ denotes calculating the angle of a complex and $(\cdot)^*$ represents the conjugation of a complex. Using the tone at $R_\text{s}/4$, the initial SPO $\tau_{0, R_\text{s} / 4}$ can be estimated by
\begin{equation}
\tau_{0, R_\text{s} / 4}=\frac{K}{\pi} \arg \left[S(N/2 - N/K/4) S^*(N/2 + N/K/4)\right].
\end{equation}
However, the estimated SPO $\tau_0$ includes the phase frequency response of the channel at $R_\text{s}/K$. When the timing error detector (TED) uses some specific frequency points, the estimated SPO using a different tone should be offset. For example, when the Godard-based TED based on frequency points near $R_\text{s}/2$ is employed, the estimated SPO using the tone at $R_\text{s}/4$ of the Y polarization requires a fixed phase offset for the TR based on Godard-based TED to achieve fast convergence, while the initial phase using the tone at $R_\text{s}/2$ of the X polarization does not require an offset. Finally, the estimated SPO initializes the timing recovery to reduce the convergence time.

\subsubsection{Fine FOE Based on Preamble A} \label{SecFFOE}
The third function of Preamble A is for fine FOE. As the frequency tones, the symbols of Preamble A are periodic in the time domain. Without loss of generality, we analyze the principle of fine FOE based on Preamble A for the X polarization $t_{\mathrm{X}}(n)$. The Preamble A with frequency offset can be modeled as
\begin{equation}
r_{\mathrm{X}}(n)=t_{{\mathrm{X}}}(n)\cdot e^{j2 \pi n \Delta f /R_\text{s}}.
\end{equation}
Then the fine FOE can be implemented by removing the phase of symbols and extracting the slope of phase when $t_{\mathrm{X}}(n)=t_{{\mathrm{X}}}(n+L)$ where $L$ is the period of the Preamble A for the X polarization. Therefore, the fine FOE can be expressed as
\begin{equation}
\begin{aligned}
\Delta f & =\frac{R_\text{s}}{2 \pi L} \cdot \arg \left[\sum_n r_{\mathrm{X}}^*(n) \cdot r_{\mathrm{X}}(n+L)\right] \\
& =\frac{R_\text{s}}{2 \pi L} \cdot \arg\left[\sum_n \left|t_{\mathrm{X}}(n)\right|^2 \cdot e^{j2 \pi L \Delta f /R_\text{s}}\right]
\end{aligned}
\end{equation}

\subsubsection{Frame Synchronization Based on Preamble B}
Frame synchronization can be implemented by calculating the timing metric based on the correlation values between transmitted and the received Preamble B. Preamble B can be expressed as $[\textbf{B}_{\mathrm{X}}~ \textbf{B}_{\mathrm{X}}~-\textbf{B}_{\mathrm{X}}]$ and $[-\textbf{B}_{\mathrm{Y}}~ \textbf{B}_{\mathrm{Y}}~\textbf{B}_{\mathrm{Y}}]$ where $\textbf{B}_{\mathrm{X}}$ and $\textbf{B}_{\mathrm{Y}}$ are two sets of CAZAC symbols with different seeds for X polarization and Y polarization, respectively. Preamble B can be used to acquire the synchronization position by seeking the prominent peak in the timing metric. We defined the correlation value $M_{\mathrm{X/Y}}(n)$ as
\begin{equation}
M_{\mathrm{X/Y}}(n) = \dfrac{2\displaystyle\sum_{i=1}^{L_\text{B}} r_{\mathrm{X/Y}}(n+i) \cdot B_{\mathrm{X/Y}}^*(n)}{\displaystyle\sum_{i=1}^{L_\text{B}}|r_{\mathrm{X/Y}}(n+i)|^2}.
\end{equation}
where $L_\text{B}$ denotes the length of $B_{\mathrm{X}}$ and $B_{\mathrm{Y}}$. To obtain the timing metric $P_{\mathrm{X/Y}}(n)$, the delay of 0, $L_\text{B}$, and $2L_\text{B}$ symbols are added to $M_{\mathrm{X/Y}}(n)$, respectively. The $M_{\mathrm{X/Y}}(n)$, $M_{\mathrm{X/Y}}(n+L_\text{B})$, and $M_{\mathrm{X/Y}}(n+2L_\text{B})$ are multiplied by the coefficients of the Preamble B, and stack their outputs over to obtain the timing metric $P_{\mathrm{X}}(n)$. In conclusion, the timing metric $P_{\mathrm{X}}(n)$ for the X polarization can be calculated by
\begin{equation}
P_{\mathrm{X}}(n)=\left|M_{\mathrm{X}}(n) + M_{\mathrm{X}}(n + L_\text{B}) - M_{\mathrm{X}}(n + 2L_\text{B}) \right|^2.
\end{equation}
The timing metric $P_{\mathrm{Y}}(n)$ for the Y polarization can be calculated by
\begin{equation}
P_{\mathrm{Y}}(n)=\left|-M_{\mathrm{Y}}(n) + M_{\mathrm{Y}}(n + L_\text{B}) + M_{\mathrm{Y}}(n + 2L_\text{B}) \right|^2.
\end{equation}
The highest peaks of $P_{\mathrm{X}}(n)$ and $P_{\mathrm{Y}}(n)$ indicate the synchronization positions of the X polarization and Y polarization, respectively.

\subsubsection{MMSE Channel Estimation Based on Preamble B}
Next, we will introduce the principle of channel estimation based on Preamble B. The mean square error (MSE) between the output of the MIMO equalizer and the transmitted signal can be expressed as
\begin{equation}
J_\mathrm{X}(\omega)=\operatorname{E}\left[\left|W_{\mathrm{XX}} R_{\mathrm{X}}+W_\mathrm{XY} R_\mathrm{Y}-T_\mathrm{X}\right|^2 \right]
\end{equation}
and
\begin{equation}
J_\mathrm{Y}(\omega)=\operatorname{E}\left[\left|W_{\mathrm{YX}} R_\mathrm{X}+W_{\mathrm{YY}} R_{\mathrm{Y}}-T_\mathrm{Y}\right|^2\right]
\end{equation}
where $\operatorname{E}(\cdot)$ is the expectation operation. $W_{\mathrm{XX}}$, $W_{\mathrm{XY}}$, $W_{\mathrm{YX}}$, and $W_{\mathrm{YY}}$ are the frequency-domain tap coefficients of the MIMO equalizer. $R_\mathrm{X/Y}$ and $T_\mathrm{X/Y}$ are the received and transmitted Preamble B of the X polarization or Y polarization in the frequency domain, respectively.

The channel estimation based on the MMSE criterion minimizes $J_\mathrm{X}(\omega)$ and $J_\mathrm{Y}(\omega)$ to obtain the tap coefficients of the MIMO equalizer. By separately taking the partial derivatives of $J_\mathrm{X}(\omega)$ for $W_{\mathrm{XX}}$ and $W_{\mathrm{XY}}$, a system of binary first-order equations can be expressed as
\begin{equation}
\operatorname{E}\left[\left(W_{\mathrm{XX}} R_{\mathrm{X}}+W_\mathrm{XY} R_\mathrm{Y}-T_\mathrm{X}\right) \cdot R_\mathrm{X}^*\right]=0
\label{dev_Jx1}
\end{equation}
and
\begin{equation}
\operatorname{E}\left[\left(W_{\mathrm{XX}} R_{\mathrm{X}}+W_\mathrm{XY} R_\mathrm{Y}-T_\mathrm{X}\right) \cdot R_\mathrm{Y}^*\right]=0,
\label{dev_Jx2}
\end{equation}
which can be solved to obtain $W_{\mathrm{XX}}$ and $W_{\mathrm{XX}}$ for minimizing $J_\mathrm{X}(\omega)$. It is worth noting that the system of binary first-order equations should have a unique solution.
Owing to the different coefficients on the elements of Preamble B, only a unique solution exists, which is proven in Appendix A. Therefore, $W_{\mathrm{XX}}$ and $W_{\mathrm{XY}}$ can be estimated by
\begin{equation}
W_{\mathrm{XX}}=\dfrac{\operatorname{E}\left(T_\mathrm{X} R_{\mathrm{X}}^*\right) \operatorname{E}\left(R_\mathrm{Y} R_\mathrm{Y}^*\right)-\operatorname{E}\left(T_\mathrm{X} R_\mathrm{Y}^*\right) \operatorname{E}\left(R_\mathrm{Y} R_{\mathrm{X}}^*\right)}{\operatorname{E}\left(R_{\mathrm{X}} R_{\mathrm{X}}^*\right) \operatorname{E}\left(R_\mathrm{Y} R_\mathrm{Y}^*\right)-\operatorname{E}\left(R_{\mathrm{X}} R_\mathrm{Y}^*\right) \operatorname{E}\left(R_\mathrm{Y} R_{\mathrm{X}}^*\right)} 
\end{equation}
and
\begin{equation}
W_{\mathrm{XY}}=\dfrac{\operatorname{E}\left(T_\mathrm{X} R_\mathrm{Y}^*\right) \operatorname{E}\left(R_{\mathrm{X}} R_{\mathrm{X}}^*\right)-\operatorname{E}\left(T_\mathrm{X} R_{\mathrm{X}}^*\right) \operatorname{E}\left(R_{\mathrm{X}} R_\mathrm{Y}^*\right)}{\operatorname{E}\left(R_{\mathrm{X}} R_{\mathrm{X}}^*\right) \operatorname{E}\left(R_\mathrm{Y} R_\mathrm{Y}^*\right)-\operatorname{E}\left(R_{\mathrm{X}} R_\mathrm{Y}^*\right) \operatorname{E}\left(R_\mathrm{Y} R_{\mathrm{X}}^*\right)}.
\end{equation}
Similarly, by separately taking the partial derivatives of $J_\mathrm{Y}(\omega)$ for $W_{\mathrm{YX}}$ and $W_{\mathrm{YY}}$, a system of binary first order equations can be expressed as
\begin{equation}
\operatorname{E}\left[\left(W_{\mathrm{YX}} R_{\mathrm{X}}+W_\mathrm{YY} R_\mathrm{Y}-T_\mathrm{Y}\right) \cdot R_\mathrm{X}^*\right]=0 
\label{dev_JY1}
\end{equation}
and
\begin{equation}
\operatorname{E}\left[\left(W_{\mathrm{YX}} R_{\mathrm{X}}+W_\mathrm{YY} R_\mathrm{Y}-T_\mathrm{Y}\right) \cdot R_\mathrm{Y}^*\right]=0.
\label{dev_JY2}
\end{equation}
which can be solved to obtain $W_{\mathrm{XX}}$ and $W_{\mathrm{XX}}$ for minimizing $J_\mathrm{Y}(\omega)$. Therefore, $W_{\mathrm{XX}}$ and $W_{\mathrm{XX}}$ can be estimated by
\begin{equation}
W_{\mathrm{YX}}=\dfrac{\operatorname{E}\left(T_\mathrm{Y} R_\mathrm{X}^*\right) \operatorname{E}\left(R_\mathrm{Y} R_{\mathrm{Y}}^*\right)-\operatorname{E}\left(T_\mathrm{Y} R_{\mathrm{Y}}^*\right) \operatorname{E}\left(R_\mathrm{Y} R_\mathrm{X}^*\right)}{\operatorname{E}\left(R_{\mathrm{X}} R_{\mathrm{X}}^*\right) \operatorname{E}\left(R_\mathrm{Y} R_\mathrm{Y}^*\right)-\operatorname{E}\left(R_{\mathrm{X}} R_\mathrm{Y}^*\right) \operatorname{E}\left(R_\mathrm{Y} R_{\mathrm{X}}^*\right)} 
\end{equation}
and
\begin{equation}
W_{\mathrm{YY}}=\dfrac{\operatorname{E}\left(T_\mathrm{Y} R_{\mathrm{Y}}^*\right) \operatorname{E}\left(R_{\mathrm{X}} R_\mathrm{X}^*\right)-\operatorname{E}\left(T_\mathrm{Y} R_\mathrm{X}^*\right) \operatorname{E}\left(R_{\mathrm{X}} R_{\mathrm{Y}}^*\right)}{\operatorname{E}\left(R_{\mathrm{X}} R_{\mathrm{X}}^*\right) \operatorname{E}\left(R_\mathrm{Y} R_\mathrm{Y}^*\right)-\operatorname{E}\left(R_{\mathrm{X}} R_\mathrm{Y}^*\right) \operatorname{E}\left(R_\mathrm{Y} R_{\mathrm{X}}^*\right)}.
\end{equation}

\begin{figure}[!t]
    \centering
    \includegraphics[width=\linewidth]{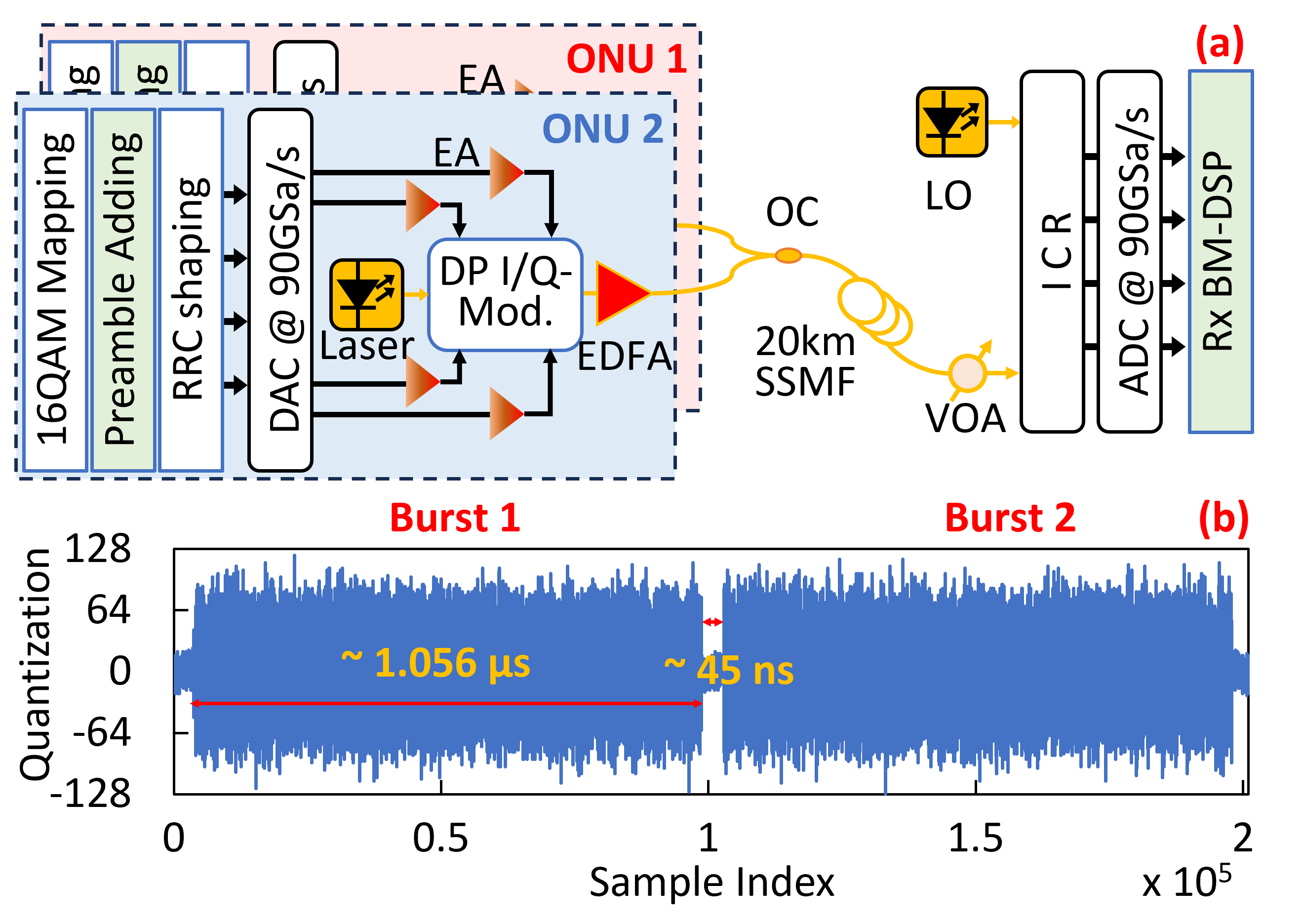}
    \caption{(a) Experimental setups of the 200G CO-TDMA with two ONUs. (b) Two received burst signals with a guard interval.}
    \label{EX setups}
\end{figure}

\subsubsection{ZF Channel Estimation Based on Preamble B}
Owing to the one-tap SOP estimation and recovery, there is almost no crosstalk between the X polarization and Y polarization, which means that the frequency responses $H_{\mathrm{YX}}$ and $H_{\mathrm{XY}}$ of the crosstalk between X polarization and Y polarization can be approximate to zeros. Therefore, the Preamble B can also be used for ZF channel estimation with less computational complexity than MMSE channel estimation. Without the consideration of noise, the Preamble B after the SOP recovery can be represented as
\begin{equation}
\begin{aligned}
\left[\begin{array}{l}
S_{\mathrm{X}} \\
S_\mathrm{Y}
\end{array}\right]&\approx \left[\begin{array}{cc}
H_{\mathrm{XX}} & 0 \\
0 & H_{\mathrm{YY}}
\end{array}\right] \times\left[\begin{array}{l}
T_\mathrm{X}  \\
T_\mathrm{Y} 
\end{array}\right] 
\end{aligned}
\end{equation}
where $H_{\mathrm{XX}}$ and $H_{YY}$ are the frequency response of the X polarization and Y polarization, respectively. The $W_{\mathrm{XX}}$ and $W_{\mathrm{YY}}$ can be calculated by
\begin{equation}\label{eq24}
\begin{aligned}
\left[\begin{array}{c}
W_{\mathrm{XX}} \\
W_{\mathrm{YY}}
\end{array}\right]&=\left[\begin{array}{c}
1/H_{\mathrm{XX}} \\
1/H_{\mathrm{YY}}
\end{array}\right] \approx \left[\begin{array}{c}
T_\mathrm{X}  / S_{\mathrm{X}} \\
T_\mathrm{Y}  / S_\mathrm{Y}
\end{array}\right]
\end{aligned}
\end{equation}
where three sets of tap coefficients can be estimated using the Preamble B, which can be averaged to obtain a more accurate set. The tap coefficients estimated by MMSE or ZF initialize the MIMO equalizer to reduce the convergence time. 

\begin{figure*}[t!]
    \centering
    \includegraphics[width=\linewidth]{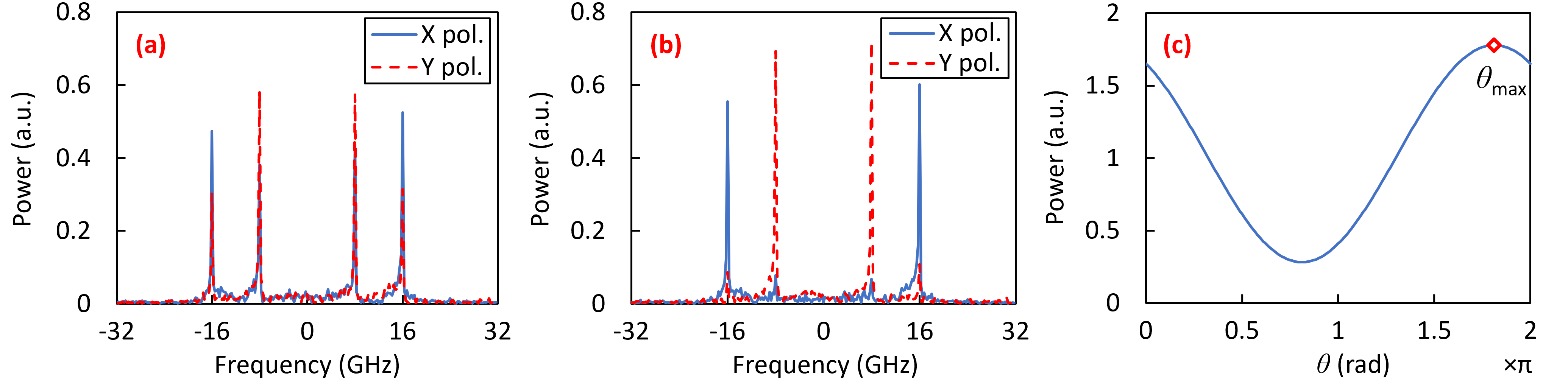}
    \caption{Spectra of the received Preamble A (a) without and (b) with the one-tap SOP estimation and recovery. (c) The power sum of two frequency tones at $R_\text{s}/2$ for the X polarization and at $R_\text{s}/4$ for the Y polarization versus $\theta$ in Eq. (\ref{Eq1}).}
    \label{SOP}
\end{figure*}

\begin{figure}[t!]
    \centering
    \includegraphics[width=0.9\linewidth]{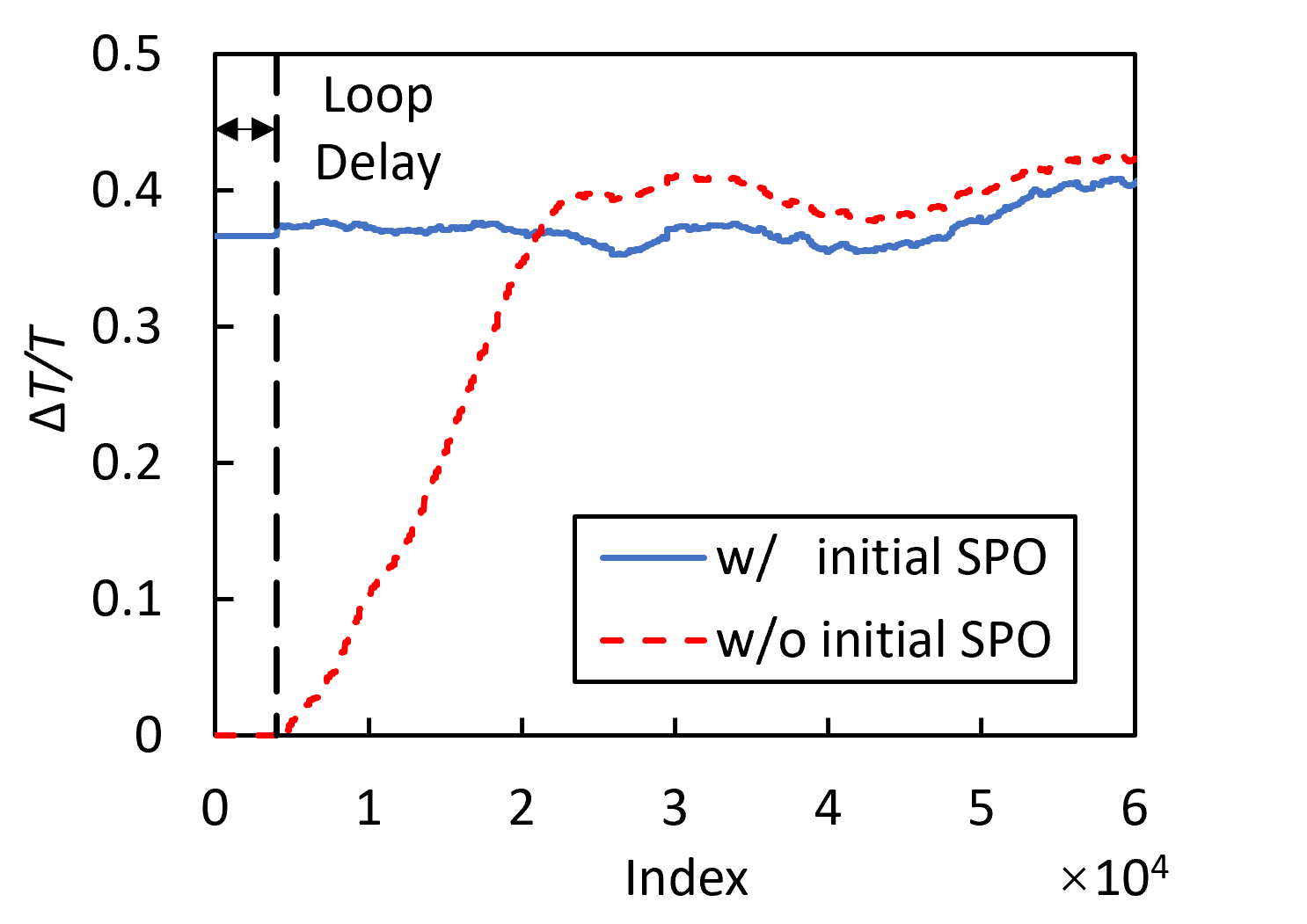}
    \caption{The SPO updating by the Godard-based feedback loop for the timing recovery with and without the initial SPO estimated by Preamble A.}
    \label{TR}
\end{figure}

\section{Experimental setups}\label{SectionIII}
An experiment of 200G CO-TDMA with two ONUs was implemented to verify the feasibility of Co-BM-DSP, shown in Fig. \ref{EX setups}(a). At the ONU, a burst signal was generated by the transmitter (Tx)-side DSP, including 16QAM mapping, preamble adding, and pulse shaping using a raised root cosine (RRC) filter with a roll-off factor of 0.1. One burst frame includes 320 preamble symbols, 32400 payload symbols, and 1046 pilot symbols. Then the digital burst signal was converted to the analog signal by a digital-to-analog converter (DAC) operating at 90GSa/s. After the electrical amplifiers (EAs), the analog signal was modulated on an optical carrier at 1550nm by a dual-polarization in-phase/quadrature modulator (DP I/Q Mod.). An external cavity laser (ECL) with a linewidth less than 100kHz was used as the source laser at the Tx. The output power of the modulator was approximately $-$12.9dBm. An Erbium-doped fiber amplifier (EDFA) boosted the optical power to approximately 3dBm. In commercial scenarios, the semiconductor optical amplifier can be employed. The amplified optical signals of the two ONUs were coupled by a 50:50 optical coupler (OC) and launched to the 20 km standard single-mode fiber (SSMF).

\begin{figure}[t!]
    \centering
    \includegraphics[width=0.9\linewidth]{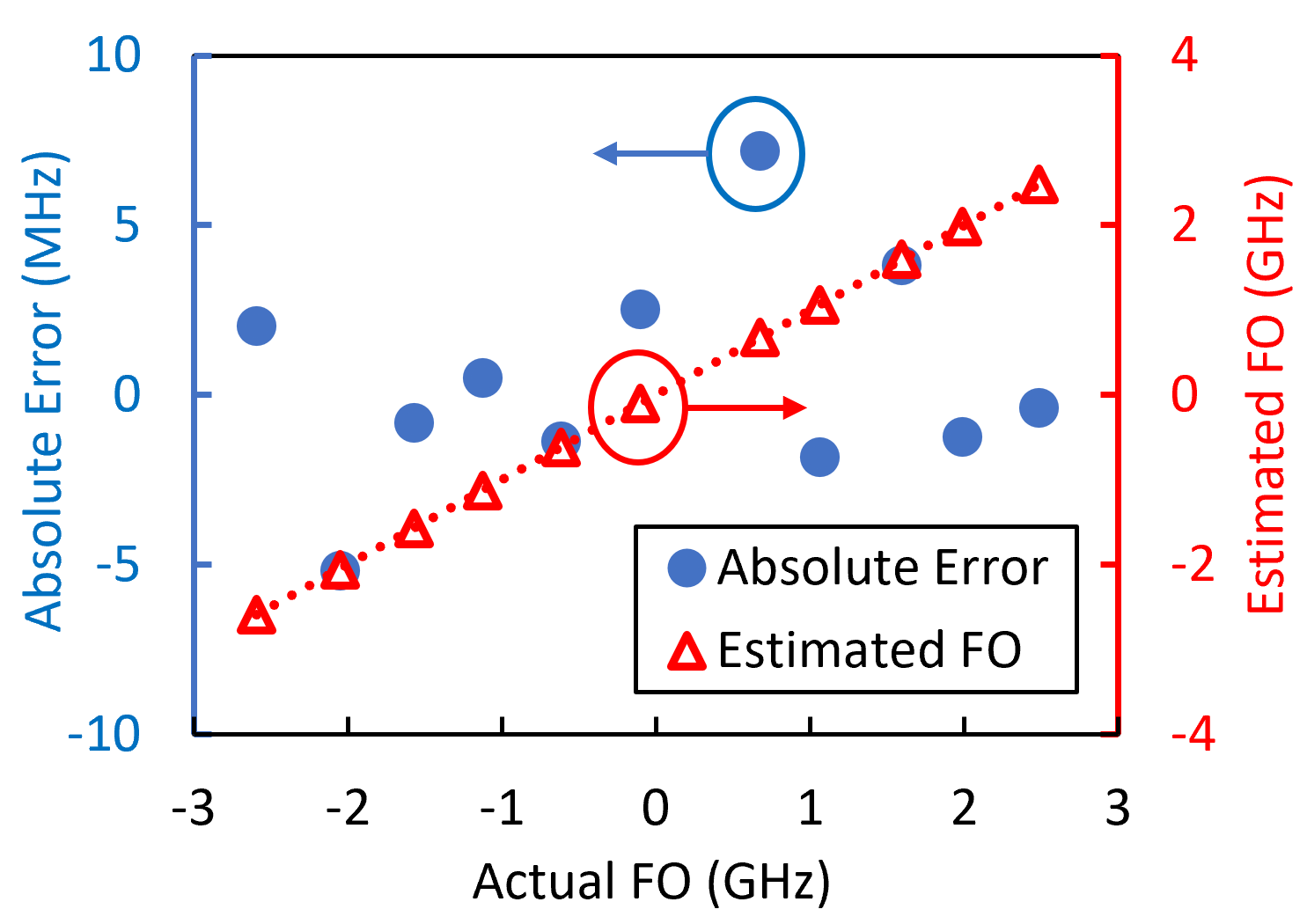}
    \caption{Absolute error between the actual and estimated frequency offset and estimated frequency offset using the Preamble A. FO: Frequency offset.}
    \label{FineFOE}
\end{figure}

At the optical line terminal (OLT), a variable optical attenuator (VOA) was used to adjust the received optical power (ROP). An integrated coherent receiver (ICR) converted the received optical signal to an analog signal. An ECL with a power of $\sim$12dBm was employed as a local oscillator (LO). The analog signal was digitized by a 90GSa/s analog-to-digital converter (ADC). Fig. \ref{EX setups}(b) shows two received burst signals with a guard interval. The duration of one burst signal was approximately 1.056 \textmu s, and the guard interval was approximately 45ns. Finally, the digital signal was recovered by the receiver (Rx)-side Co-BM-DSP as shown in Fig. \ref{DSP}.

\begin{figure}[t!]
    \centering
    \includegraphics[width=\linewidth]{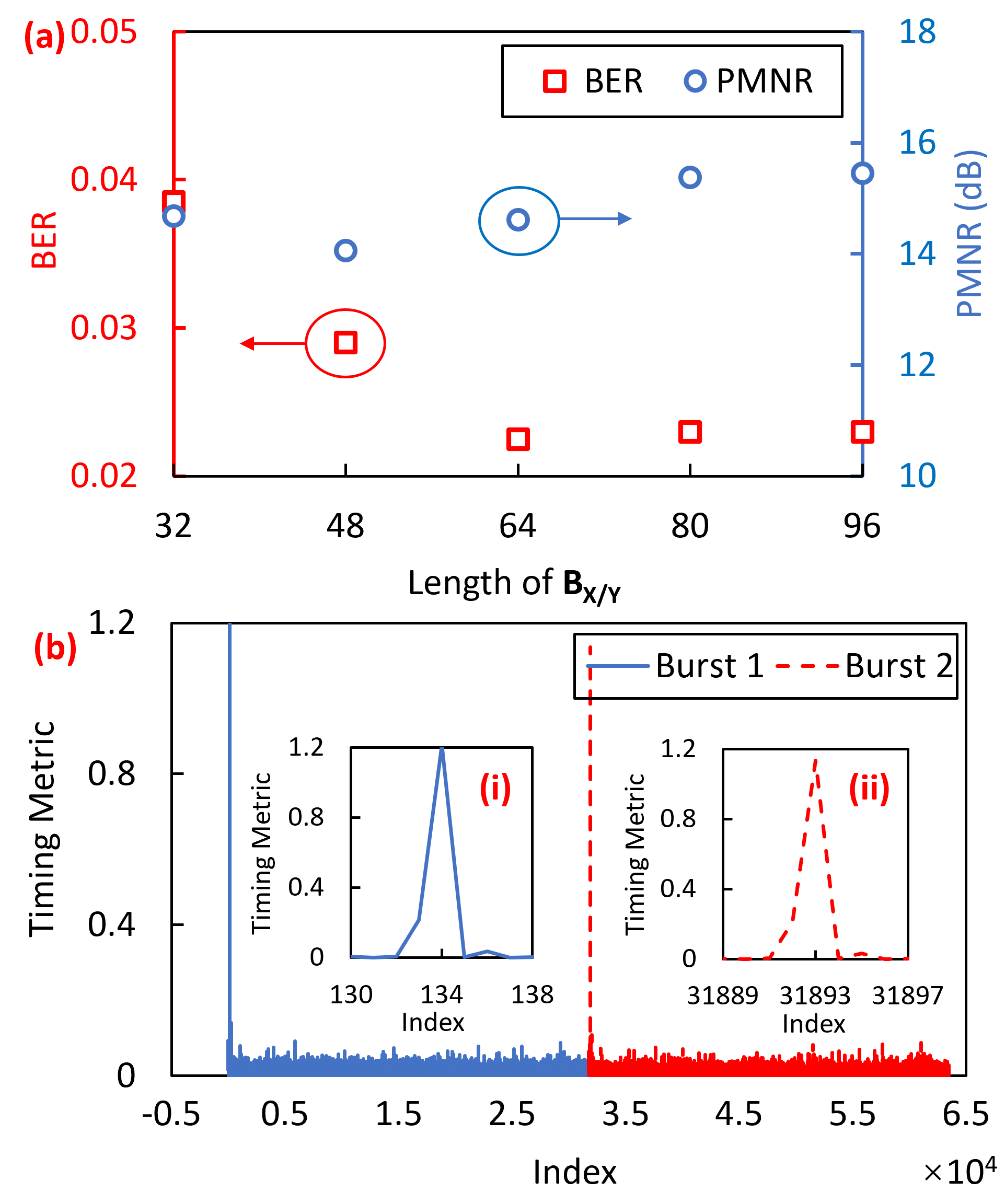}
    \caption{(a) BER of the first one thousand bits and PMNR of the synchronization peak versus the length of $\textbf{B}_{\mathrm{X/Y}}$. (b) Timing metric of the synchronization peaks using the Preamble B for Burst 1 and Burst 2 signals. Insets are zoomed timing metrics of the (i) Burst 1 and (ii) Burst 2 around the peaks.}
    \label{SYNC}
\end{figure}

\section{Experimental Results}\label{SectionIV}
Figures \ref{SOP}(a) and (b) show the spectra of the received Preamble A without and with the one-tap SOP estimation and recovery, respectively. Before the one-tap SOP recovery, the $R_\text{s}/2$ and $R_\text{s}/4$ tones interfere between the X polarization and Y polarization. After the one-tap SOP estimation and recovery, most power of the $R_\text{s}/2$ tones appears at the X polarization, and most power of the $R_\text{s}/4$ tones appears at the Y polarization. Fig. \ref{SOP}(c) shows the power sum of two frequency tones at $R_\text{s}/2$ for the X polarization and at $R_\text{s}/4$ for the Y polarization versus $\theta$ in Eq. (\ref{Eq1}). With the parameter $\theta$ in the range from 0 to $2\pi$, the power sum is a sine function. Therefore, there is only one $\theta_{\text{max}}$ to maximize the sine function, which is the estimated relative phase difference between the X polarization and Y polarization.

Preamble A can also implement the SPO estimation based on the feed-forward algorithm in Section \ref{SecTR}. Fig. \ref{TR} shows the SPO updating by the Godard-based feedback loop for the timing recovery with and without the initial SPO estimated by Preamble A. Assuming the loop delay is approximately 2000 symbols (i.e. 100 symbols/beat$\times$20 beats), the SPO update does not work within the first 2000 symbols. The timing recovery with the initial SPO can converge faster than that without the initial SPO. Meanwhile, the more accurate initial SPO means better performance. The fine FOE can be realized by Preamble A based on the principle in Section \ref{SecFFOE}. Fig. \ref{FineFOE} shows the absolute error between the actual and estimated frequency offset by the fine FOE based on the Preamble A. The fine FOE can achieve an absolute error within approximately 10MHz when the added frequency offsets are from $-$3GHz to 3GHz. In conclusion, the designed Preamble A and algorithms can accurately estimate the SOP, SPO, and frequency offset.

Figure \ref{SYNC}(a) shows the bit error ratio (BER) and peak-to-maximum-noise ratio (PMNR) of the synchronization peak versus the length of $\textbf{B}_{\mathrm{X/Y}}$ at the ROP of $-$32dBm. As the length of $\textbf{B}_{\mathrm{X/Y}}$ increases, the channel estimation becomes more accurate and BER performance improves. When the length of $\textbf{B}_{\mathrm{X/Y}}$ is greater than or equal to 64, the BER performance is stable. The length of the $\textbf{B}_{\mathrm{X/Y}}$ can be set to 64 to achieve a BER under the 20\%-overhead forward error correction (FEC) limit of 2.4$\text{e}^{-2}$. Meanwhile, the PMNR is approximately 15dB when the length of the $\textbf{B}_{\mathrm{X/Y}}$ is set to 64. Fig. \ref{SYNC}(b) depicts the timing metric of synchronization peaks using the Preamble B for Burst 1 and Burst 2. The length of $\textbf{B}_{\mathrm{X/Y}}$ was set to 64. The power of synchronization peaks for Burst 1 and Burst 2 are much higher than the noise. There is only one peak for each burst signal as Insets (i) and (ii) show. Therefore, accurate synchronization positions can be obtained by frame synchronization based on the Preamble B.

\begin{figure}[t!]
    \centering
    \includegraphics[width=\linewidth]{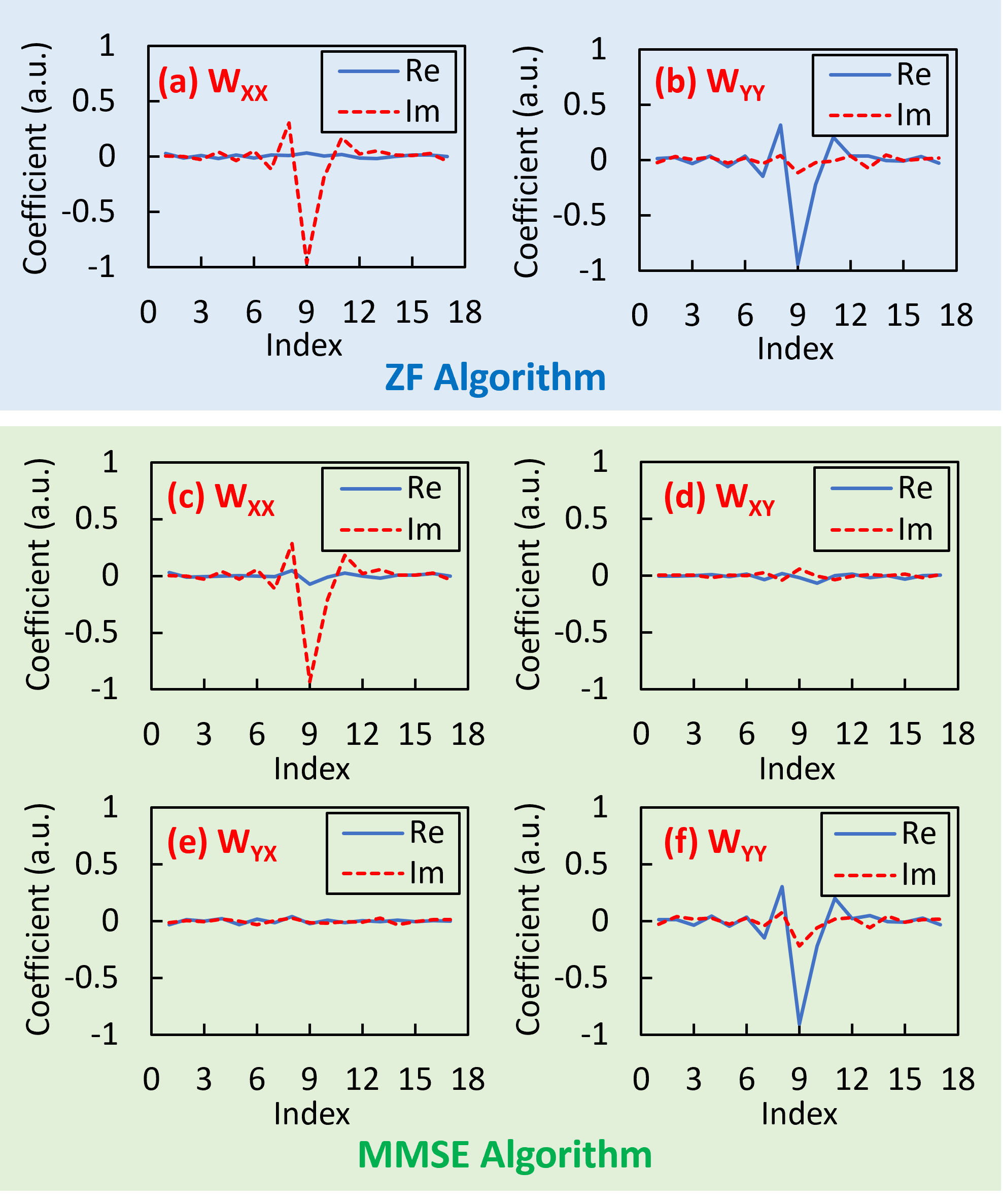}
    \caption{The real (Re) part and imaginary (Im) part of the estimated tap coefficients (a) $\mathbf{W}_{\text{XX}}$ and (b) $\mathbf{W}_{\text{YY}}$ using ZF algorithm. The Re part and Im part of the estimated tap coefficients (c) $\mathbf{W}_{\text{XX}}$, (d) $\mathbf{W}_{\text{XY}}$, (e) $\mathbf{W}_{\text{YX}}$, and (f) $\mathbf{W}_{\text{YY}}$ using MMSE algorithm.}
    \label{H}
\end{figure}

When SOP has been recovered by using Preamble A, the tap coefficients $\mathbf{W}_{\text{XX}}$ and $\mathbf{W}_{\text{YY}}$ for the MIMO equalizer can be estimated by the ZF algorithm based on Preamble B, which are shown in Fig. \ref{H}(a) and Fig. \ref{H}(b), respectively. Since it is assumed that there is no crosstalk between the X polarization and Y polarization, the tap coefficients $\mathbf{W}_{\text{YX}}$ and $\mathbf{W}_{\text{XY}}$ are zeros and not shown. Fig. \ref{H}(c), Fig. \ref{H}(d), Fig. \ref{H}(e), and Fig. \ref{H}(f) show the tap coefficients $\mathbf{W}_{\text{XX}}$, $\mathbf{W}_{\text{XY}}$, $\mathbf{W}_{\text{YX}}$, and $\mathbf{W}_{\text{XX}}$ for the MIMO equalizer estimated by the MMSE algorithm using Preamble B, respectively. The tap coefficients $\mathbf{W}_{\text{XY}}$ and $\mathbf{W}_{\text{YX}}$ are close to 0, which denotes that the SOP has been estimated and recovered effectively based on Preamble A. The tap coefficients $\mathbf{W}_{\text{XX}}$ and $\mathbf{W}_{\text{YY}}$ estimated by the ZF algorithm are similar to those estimated by the MMSE algorithm. 

\begin{figure}[t!]
    \centering
    \includegraphics[width=\linewidth]{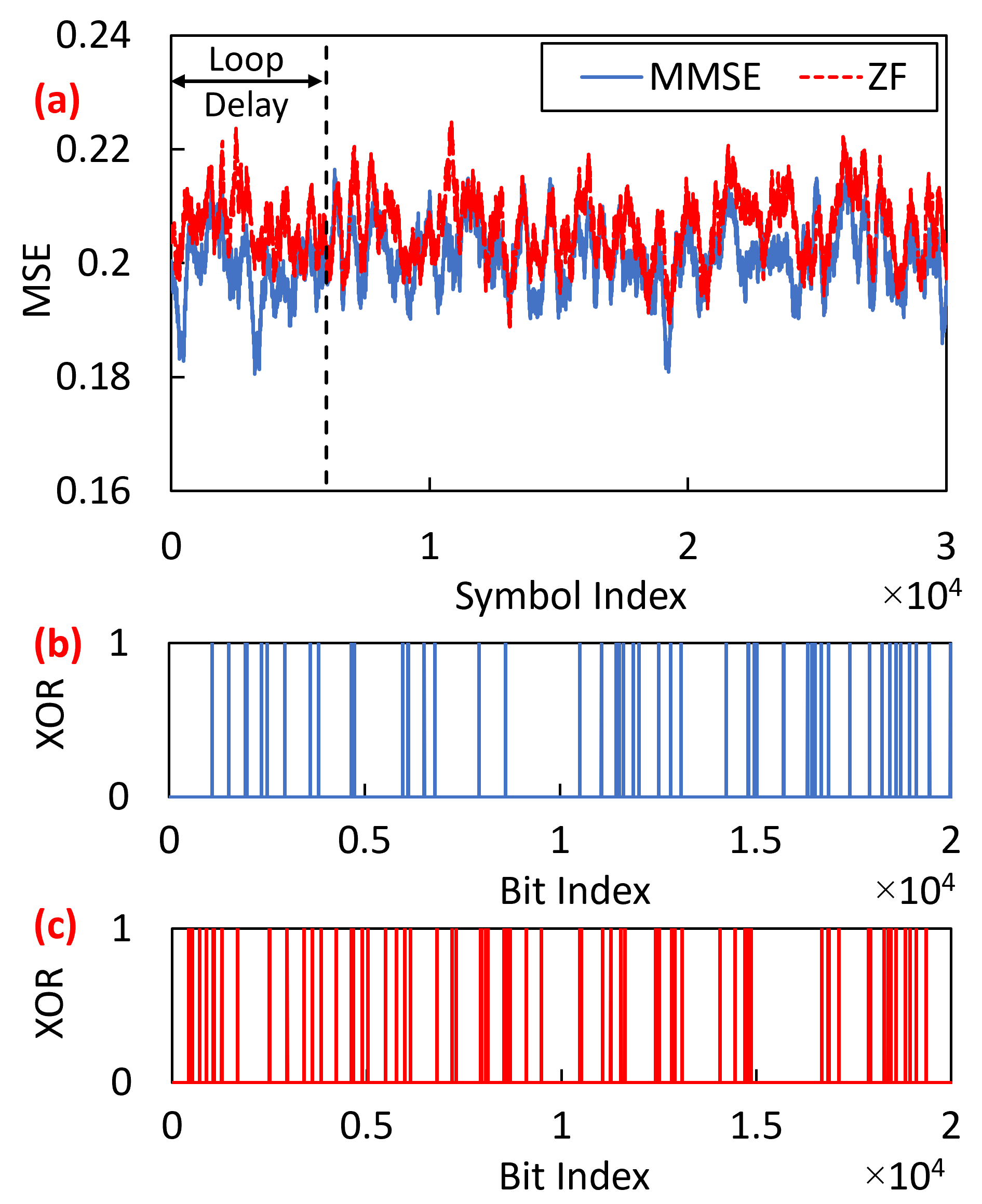}
    \caption{(a) MSE curve of the MIMO equalizer with the initialized tap coefficients estimated by MMSE and ZF algorithms based on the Preamble B. The error distributions of the first 20000 bits using tap coefficients estimated by the (b) MMSE and (c) ZF algorithms. }
    \label{MSE}
\end{figure}

Figure \ref{MSE} depicts the mean square error (MSE) curve of the MIMO equalizer with the initialized tap coefficients estimated by MMSE and ZF algorithms based on Preamble B. After the initialization of tap coefficients, the DD-LMS algorithm updates the tap coefficients. In previous works, the ZF algorithm usually requires a preamble with half zero-padding symbols, doubling the length of the preambles. Owing to the SOP estimation and recovery using Preamble A, the ZF algorithm can be realized using Preamble B, which has a low computational complexity. However, the MSE of the ZF algorithm is still higher than that of the MMSE algorithm. Assuming the loop delay of the MIMO equalizer with the DD-LMS algorithm is approximately 6000 symbols (i.e. 100 symbols/beat$\times$60 beats), the updating for the tap coefficients does not work. Therefore, the more accurate initialized tap coefficients mean better performance at the first 6000 symbols (i.e. $\sim$20000 bits). Figs. \ref{MSE}(b) and (c) show the error distributions of the first 20000 bits using the tap coefficients estimated by the MMSE and ZF algorithms at the ROP of $-$27dBm, respectively. The error distribution using the MMSE algorithm is sparser than that using the ZF algorithm. In conclusion, the MMSE algorithm can estimate the tap coefficients more accurately but has higher computational complexity compared to the ZF algorithm.

\begin{figure}[t!]
    \centering
    \includegraphics[width=\linewidth]{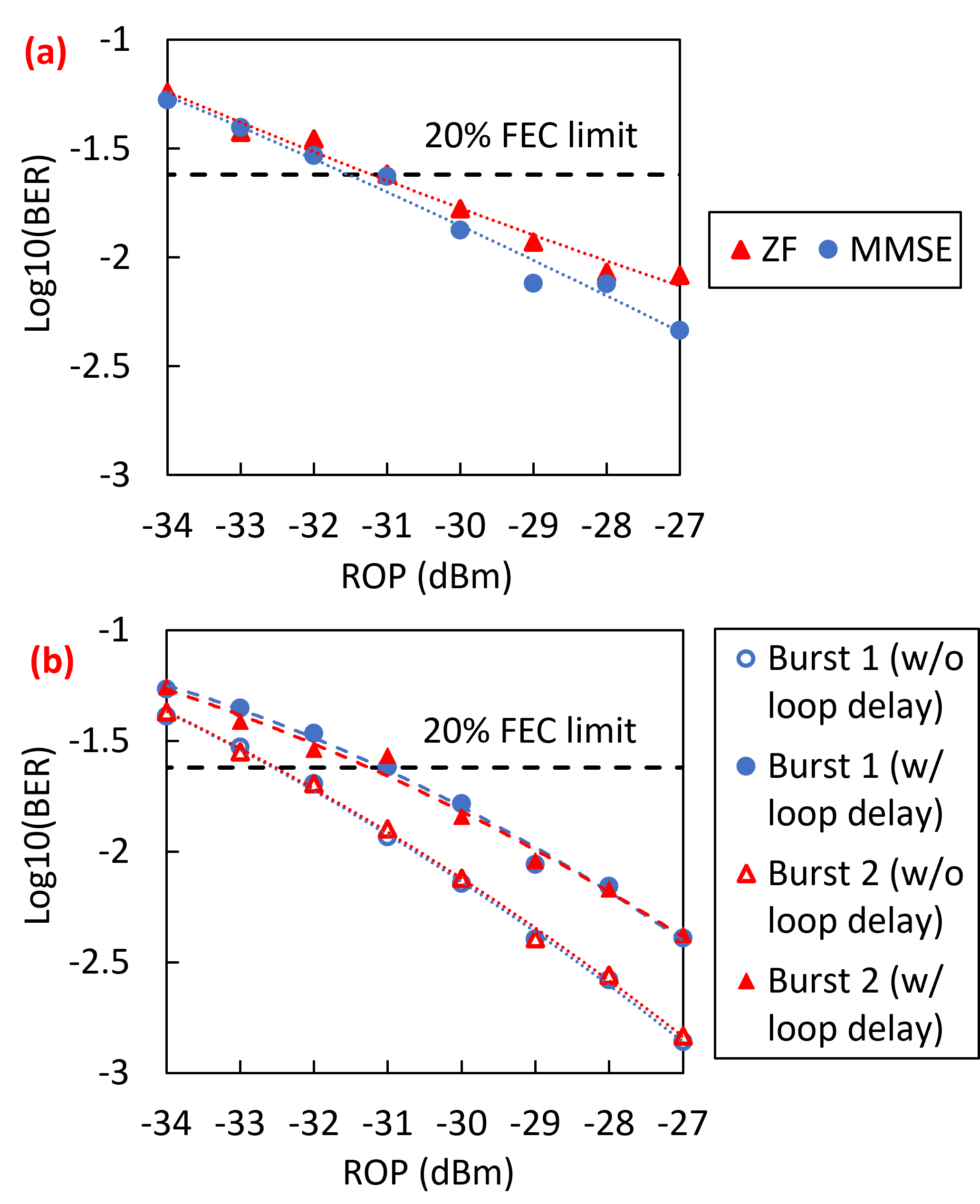}
    \caption{(a) BER versus ROP for the first 20000 bits when the MMSE and ZF algorithms are employed, and the DD-LMS algorithm does not work. (b) The BER versus ROP of Burst 1 and Burst 2 without and with the loop delay when the DD-LMS algorithm works.}
    \label{BER}
\end{figure}

Fig. \ref{BER}(a) shows the BER versus ROP for the first 20000 bits when the MMSE and ZF algorithms are employed, and the DD-LMS algorithm does not work. At the high ROP, the MMSE algorithm has a clear performance gain compared to the ZF algorithm. At the 20\% FEC limit, MMSE and ZF algorithms have similar performance. Therefore, owing to the SOP estimation and recovery using Preamble A, the low-complexity ZF algorithm based on Preamble B can potentially initialize the tap coefficients. Fig. \ref{BER}(b) shows BER versus ROP of Burst 1 and Burst 2 without and with the loop delay when the DD-LMS algorithm works. The 1.5dB higher ROP is required when the loop delay is considered. Therefore, the loop delay must be considered in the confirmatory experiment for the performance verification of the MIMO equalizer. The ROPs of Burst 1 and Burst 2 are both $-$31 dBm when the loop delay is considered. For the uplink of 200G CO-TDMA, the optical power budget can achieve 34 dB, which satisfies the requirement of commercial optical access.

\section{Conclusions}\label{SectionV}
This paper proposes a Co-BM-DSP based on an approximately 10ns designed preamble for upstream reception of 200G CO-TDMA using 32GBaud PDM-16QAM. The designed preamble can be decomposed into a preamble of frequency tones with 128 symbols and a preamble of 192 CAZAC symbols. The frequency tones can be used for SOP, frequency offset, and SPO estimation. The CAZAC symbols can be employed for estimating frame synchronization and tap coefficients of the MIMO equalizer. Based on the designed preamble, the SPO of timing recovery and tap coefficients of the MIMO equalizer can be estimated quickly by the feed-forward algorithms and initialized to put the feedback algorithms under the right status. It is worth noting that our offline experiments in the Lab are relatively ideal. The preamble length can be properly increased to ensure the performance of the actual scenarios. In conclusion, the proposed Co-BM-DSP with a compact preamble makes it more possible to be applied in the future CO-TDMA.

\appendices
\section{Proof of The Unique Solutions Using MMSE Channel Estimation Based on Preamble B}\label{AppendicesA}

This appendix proves that there are unique solutions for Eq. (\ref{dev_Jx1}), Eq. (\ref{dev_Jx2}), Eq. (\ref{dev_JY1}), and Eq. (\ref{dev_JY2}), which denotes the MMSE channel estimation based on the Preamble B. The Preamble B consists of three sets of 64 CAZAC symbols, which can be expressed as $\mathbf{T}_{\text{X1}}$, $\mathbf{T}_{\text{X2}}$, and $\mathbf{T}_{\text{X3}}$ for the X polarization and $\mathbf{T}_{\text{Y1}}$, $\mathbf{T}_{\text{Y2}}$, and $\mathbf{T}_{\text{Y3}}$ for the Y polarization, respectively. At the receiver end, the received three sets of 64 CAZAC symbols are expressed as $\mathbf{R}_{\text{X1}}$, $\mathbf{R}_{\text{X2}}$, and $\mathbf{R}_{\text{X3}}$ for the X polarization and $\mathbf{R}_{\text{Y1}}$, $\mathbf{R}_{\text{Y2}}$, and $\mathbf{R}_{\text{Y3}}$ for the Y polarization, respectively. Therefore, Eq. (\ref{dev_Jx1}) and Eq. (\ref{dev_Jx2}) can be extended as Eq. (\ref{app_eq1.1}) and Eq. (\ref{app_eq1.2}). Moreover, the Eq. (\ref{app_eq1.1}) and Eq. (\ref{app_eq1.2}) can be rewritten as 
\newcounter{TempEqCnt3}
\setcounter{TempEqCnt3}{\value{equation}}
\setcounter{equation}{26}
\begin{equation}
\sum_{i=1}^3R_{\text{X} i} R_{\text{X} i}^* \cdot W_{\mathrm{XX}}+\sum_{i=1}^3 R_{\text{Y} i}R_{\text{X} i}^* \cdot W_{\mathrm{XY}}=\sum_{i=1}^3 R_{\text{X} i}^*T_{\text{X} i}
\label{app_eq2.1}
\end{equation}
and
\begin{equation}
\sum_{i=1}^3R_{\text{X} i} R_{\text{Y} i}^* \cdot W_{\mathrm{XX}}+\sum_{i=1}^3 R_{\text{Y} i}R_{\text{Y} i}^* \cdot W_{\mathrm{XY}}=\sum_{i=1}^3 R_{\text{Y} i}^*T_{\text{X} i}.
\label{app_eq2.2}
\end{equation}
Typically, there are three cases for solving Eq. (\ref{app_eq2.1}) and Eq. (\ref{app_eq2.2}) to obtain $\mathbf{W}_{\text{XX}}$ and $\mathbf{W}_{\text{XY}}$, including unique and infinite, and no solutions.
\newcounter{TempEqCnt1}
\setcounter{TempEqCnt1}{\value{equation}}
\setcounter{equation}{24}
\begin{figure*}
\begin{equation}
\frac{1}{3}\sum_{i=1}^3\left(W_{\mathrm{XY}} R_{\text{X} i}+W_{\mathrm{XY}} R_{\text{Y} i}-T_{\text{X} i}\right) \cdot R_{\text{X} i}^*=\frac{1}{3}\left(\sum_{i=1}^3R_{\text{X} i} R_{\text{X} i}^* \cdot W_{\mathrm{XX}}+\sum_{i=1}^3 R_{\text{Y} i}R_{\text{X} i}^* \cdot W_{\mathrm{XY}}-\sum_{i=1}^3 R_{\text{X} i}^*T_{\text{X} i}\right)=0
\label{app_eq1.1}
\end{equation}
\hrulefill
\end{figure*}
\begin{figure*}
\begin{equation}
\frac{1}{3}\sum_{i=1}^3\left(W_{\mathrm{XY}} R_{\text{X} i}+W_{\mathrm{XX}} R_{\text{Y} i}-T_{\text{X} i}\right) \cdot R_{\text{Y} i}^*=\frac{1}{3}\left(\sum_{i=1}^3R_{\text{X} i} R_{\text{Y} i}^* \cdot W_{\mathrm{XX}}+\sum_{i=1}^3 R_{\text{Y} i}R_{\text{Y} i}^* \cdot W_{\mathrm{XY}}-\sum_{i=1}^3 R_{Y i}^*T_{\text{X} i}\right)=0
\label{app_eq1.2}
\end{equation}
\hrulefill
\end{figure*}
\begin{figure*}
\newcounter{TempEqCnt2}
\setcounter{TempEqCnt2}{\value{equation}}
\setcounter{equation}{28}
\begin{equation}
\left|\begin{array}{ll}
\sum_{i=1}^3 R_{\text{X} i}R_{\text{X} i}^* & \sum_{i=1}^3 R_{\text{Y} i}   R_{\text{X} i}^* \\
\sum_{i=1}^3 R_{\text{X} i}R_{\text{Y} i}^* & \sum_{i=1}^3 R_{\text{Y} i}   R_{\text{Y} i}^*
\end{array}\right|=\left|R_{\text{X} 1}   R_{\text{Y} 2}-R_{\text{Y} 1}   R_{\text{X} 2}\right|^2+\left|R_{\text{X} 1}   R_{\text{Y} 3}-R_{\text{Y} 1}   R_{\text{X} 3}\right|^2+\left|R_{\text{X} 2}   R_{\text{Y} 3}-R_{\text{Y} 2}   R_{\text{X} 3}\right|^2 \geq 0
\label{app_eq3}
\end{equation}
\hrulefill
\end{figure*}

From Eq. (\ref{app_eq2.1}) and Eq. (\ref{app_eq2.2}), the determinant of their coefficients can be expressed as Eq. (\ref{app_eq3}), which is great than or equal to zero. If there is a nonzero item in Eq. (\ref{app_eq3}) due to 
\begin{equation}
R_{\mathrm{X}i}R_{\mathrm{Y}j} \neq R_{\mathrm{Y}i}R_{\mathrm{X}j}, 
\end{equation}
the determinant is not equal to zero where $i\neq j$ are both from 1 to 3. Therefore, Eq. (\ref{app_eq2.1}) and Eq. (\ref{app_eq2.2}) have a unique solution. If all the items are equal to zero in Eq. (\ref{app_eq3}) due to
\begin{equation}
R_{\mathrm{X}i}R_{\mathrm{Y}j} =R_{\mathrm{Y}i}R_{\mathrm{X}j},
\label{app_eq4}
\end{equation}
the determinant is equal to zero. Eq. (\ref{app_eq2.1}) and Eq. (\ref{app_eq2.2}) may have infinite solutions or no solution. To confirm which one is right, we should consider
\begin{equation}
\begin{aligned}
\sum_{i=1}^3 & R_{\text{X} i} R_{\text{X} i}^*\sum_{i=1}^3 R_{\text{Y} i}^*T_{\text{X} i}-\sum_{i=1}^3 R_{\text{X} i}^* T_{\text{X} i} \sum_{i=1}^3 R_{\text{X} i} R_{\text{Y} i}^* \\
= &\underbrace{\left(R_{\text{X} 1} R_{\text{Y} 2}-R_{\text{Y} 1} R_{\text{X} 2}\right)^*}_0 \left(R_{\text{X} 1} T_{\text{X} 2}-T_{\text{X} 1} R_{\text{X} 2}\right)+ \\
& \underbrace{\left(R_{\text{X} 1} R_{\text{Y} 3}-R_{\text{Y} 1} R_{\text{X} 3}\right)^*}_0 \left(R_{\text{X} 1} T_{\text{X} 3}-T_{\text{X} 1} R_{\text{X} 3}\right)+ \\
& \underbrace{\left(R_{\text{X} 2} R_{\text{Y} 3}-R_{\text{Y} 2} R_{\text{X} 3}\right)^*}_0 \left(R_{\text{X} 2} T_{\text{X} 3}-T_{\text{X} 2} R_{\text{X} 3}\right),
\end{aligned}
\label{app_eq5}
\end{equation}
which is equal to zero. Therefore, there are infinite solutions for Eq. (\ref{app_eq2.1}) and Eq. (\ref{app_eq2.2}). In conclusion, there are only two cases for solving Eq. (\ref{app_eq2.1}) and Eq. (\ref{app_eq2.2}), including unique and infinite solutions.

Using the designed Preamble B, the determinant is not equal to zero due to $R_{\mathrm{X}1}R_{\mathrm{Y}2} \neq R_{\mathrm{Y}1}R_{\mathrm{X}2}$ and $R_{\mathrm{X}3}R_{\mathrm{Y}2} \neq R_{\mathrm{Y}3}R_{\mathrm{X}2}$. Therefore, the Eq. (\ref{app_eq2.1}) and Eq. (\ref{app_eq2.2}) have a unique solution. However, without the coefficients of $[1, 1, -1]$ for the X polarization and $[-1, 1, 1]$ for the Y polarization,  the determinant and Eq. (\ref{app_eq5}) are equal to zero, which would lead to infinite solutions of $\mathbf{W}_{\text{XX}}$ and $\mathbf{W}_{\text{XY}}$ for Eq. (\ref{app_eq2.1}) and Eq. (\ref{app_eq2.2}). A similar proof can be applied to Eq. (\ref{dev_JY1}) and Eq. (\ref{dev_JY2})  to solving the $\mathbf{W}_{\text{YX}}$ and $\mathbf{W}_{\text{YY}}$. In conclusion, MMSE channel estimation based on the designed Preamble B is theoretically proven to have unique solutions for obtaining the tap coefficients, ensuring the reliability of the Co-BM-DSP. 

\bibliographystyle{ieeetr}
\bibliography{Ref}
\end{document}